\documentclass[twocolumn,11pt]{article}

\usepackage{amsmath}
\usepackage{amssymb}
\usepackage{graphicx}
\usepackage{chicago}

\newcommand{\mat}[1]{{\mbox{\bfseries \rmfamily#1}}}

\begin{document}

\title{Finite genome size can halt Muller's ratchet}

\author{Tor Schoenmeyr and Claus O. Wilke\\\mbox{}\\
  \it Digital Life Laboratory\\\it
  California Institute of Technology, Mail-Code 136-93\\\it Pasadena, CA
  91125}

\date{}
\maketitle

{\bf Abstract:} We study the accumulation of deleterious mutations in a
haploid, asexually reproducing population, using analytical models and
computer simulations. We find that Muller's ratchet can come to a halt in
small populations as a consequence of a finite genome size only, in the
complete absence of backward or compensatory mutations, epistasis, or
recombination. The origin of this effect lies in the fact that the number of
loci at which mutations can create considerable damage decreases with every
turn of the ratchet, while the total number of mutations per genome and
generation remains constant. Whether the ratchet will come to a halt
eventually depends on the ratio of the \emph{per-locus} deleterious mutation
rate $u$ and the selection strength $s$. For sufficiently small $u/s$, the
ratchet halts after only a few clicks. We discuss the implications of our
results for
bacterial and virus evolution.\\


\noindent
A major evolutionary pressure that distinguishes finite asexual from sexual
populations experience is Muller's ratchet, i.e., the continual loss of those
individuals from the population that carry the smallest load of mutations
\shortcite{Muller64,Felsenstein74}. Sexual populations can regenerate
individuals with reduced number of mutations through recombination, while
asexual populations cannot. This may be one of the main advantages of sexual
recombination
\shortcite{Felsenstein74,MaynardSmith78,Takahata82,Pamiloetal87,AntezanaHudson97}.
Besides its importance for theories of the evolution and maintainance of sex,
Muller's ratchet may also play a significant role in bacterial or viral
dynamics, where host to host transmission often creates severe bottlenecks
\shortcite{Chao90,AnderssonHughes96}.  Conventional wisdom says that Muller's
ratchet proceeds at a constant rate, as a result of which an asexual
population may eventually go extinct \shortcite{LynchGabriel90,Gabrieletal93}.
The main assumptions that enter the classic ratchet model are a vanishing
probability of back mutations, a genome with an infinite number of loci, and
mutations that do not interact.  When epistatic interactions are taken into
account, Muller's ratchet will slow down over time
\shortcite{Charlesworthetal93,Kondrashov94}, though it will not necessarily
come to a halt \cite{Butcher95}. On the other hand, if back mutations are
assumed to be frequent and the genome is finite, then the ratcheting process
will always stop eventually, even in the absence of epistatic interactions
\shortcite{WoodcockHiggs96,PruegelBennet97}. A similar result can be found if
there is sufficient supply of compensatory mutations~\cite{WagnerGabriel90}.
However, since the majority of studies focuses on the classic model with
infinite genome
\shortcite{Haigh78,Pamiloetal87,Stephanetal93,HiggsWoodcock95,Gessler95,Charlesworth97,GordoCharlesworth2000a,GordoCharlesworth2000b},
it is not entirely clear how Muller's ratchet affects bacteria or viruses,
which contain at most a couple of hundred genes.

Here, we investigate the selection-mutation balance for a finite number of
loci, and we also consider arbitrary forward and
back mutation rates.
Perhaps not so surprisingly, we find that if the back mutation rate is small
but non-zero, selection-mutation balance stabilizes the population before the
genome is completely deteriorated. More importantly, however, the assumption
of a finite number of loci alone guarantees that even in the complete absence
of back mutations, Muller's ratchet can come to a complete halt within a
biologically relevant time frame if selection is sufficiently strong.

\begin{center}
{\sc Infinite population size}
\end{center}

As in the case of an infinitely long genome, the deterministic
mutation-selection balance of an infinite population is a useful basis for the
subsequent analysis of finite population effects \shortcite{Haigh78}. We
therefore give a brief treatment of the infinite population case, and discuss
the influence of various rates of back mutations on the equilibrium
distribution of deleterious mutations.

The deterministic mutation-selection equilibrium can be calculated
straightforwardly from the eigenvector corresponding to the largest eigenvalue
of the transition matrix $\mat W$ \shortcite{Eigenetal88}. The transition
matrix contains the rates at which genotypes are produced as offspring of
other genotypes, i.e., an entry $w_{ij}$ in the transition matrix is given by
the fitness of genotype $j$ times the probability that genotype $j$ mutates
into genotype $i$. In the present case, the analysis can be simplified
because we disregard epistatic interactions, in which case it is
sufficient to calculate the equilibrium point for a single locus. The
generalization to a finite number of loci $L$ is trivial \cite{Rumschitzki87}.

We assume a single locus with two alleles, '$+$' and '$-$'. The '$+$' allele
carries a fitness of 1, and the '$-$' allele carries a fitness of $1-s$.
Mutations from '$+$' to '$-$' may occur with a rate $u$ per locus. We call
these mutations forward (or deleterious) mutations. Conversely, mutations from
'$-$' to '$+$' occur with a rate $v$, and we call these mutations back
mutations. The transition matrix for this single-locus, two-allele model is
given by
\begin{equation}
  \mat W = \begin{pmatrix} 1-u & (1-s) v \\ u & (1-s)(1-v) \end{pmatrix}\,.
\end{equation}
From diagonalization, we find that the '$-$' allele is present in the population
at a concentration of
\begin{equation}
  a = \frac{1}{2}\Big[1-v+\frac{u+v}{s}-\sqrt{\Big(1-v+\frac{u+v}{s}\Big)^2-\frac{4u}{s}}\Big]\,.
\end{equation}
The concentration of the '$+$' allele is correspondingly $1-a$. In the case of
$L$ loci, this result generalizes to
\begin{equation}\label{det-concs}
  x_k=\binom{L}{k} a^k(1-a)^{L-k}\,,
\end{equation}
where $x_k$ is the concentration of genomes that carry $k$ alleles of the
'$-$' type (we will in the following simply say that these genomes carry $k$
mutations). From \eqref{det-concs}, the average number of mutations $\langle
k\rangle$ in the population follows as $\langle k\rangle=aL$.

If we set the back mutation rate to $v=0$, we find the simple expression
\begin{equation}
\frac{\langle k\rangle}{L} = a = \left\{\begin{array}{cl} 1 & \mbox{for $u>s$} \\
                u/s & \mbox{for $u\leq s$.} \end{array}\right.
\end{equation}
We learn that the fate of an infinite population without back mutations is
determined by the ratio between the rate of deleterious mutations per locus
$u$ and the selection strength $s$. When $u$ is of the order of $s$ or larger,
then $\langle k\rangle=L$, which means that the population will eventually
consist exclusively of individuals that have been hit by mutations at all
loci. For $u$ smaller than $s$, on the other hand, the majority of individuals
in the population will accumulate only a small number of mutations.

In Fig.~\ref{fig:phase_trans}, we show the average fraction of
mutated loci in the population, $\langle k\rangle/L$, as a function of the
forward mutation rate $u$ and for various back mutation rates $v$.  Note the
sharpness of the transition at $u/s=1$. If $u$ is only a factor of 10 smaller
than $s$, then on average each individual carries mutations in only 10\% of
the loci. We observe further that the ratio of $u/s$ plays a more important
role in determining the equilibrium point of the population than the back
mutation rate $v$. Unless $v$ is of the order of $u$, the deviations from the
case of $v=0$ are small.

It is worth mentioning that if we substitute $a=u/s$ into \eqref{det-concs}
and take the limit $L\rightarrow\infty$ while keeping
the genomic mutation rate $U=uL$ constant, we recover the Poisson
distribution of mutants that is normally assumed in models of Muller's
ratchet with infinite genome size \shortcite{Haigh78}.

\begin{center}
{\sc Finite population size}
\end{center}

\noindent
\textbf{Simulation methods}

\noindent
All simulation results reported in this work were obtained in the following
manner. We keep track of $L+1$ variables $n_i$, each of which counts the
number of individuals with $i$ mutations currently present in the population.
In order to propagate the population through one round of selection and
mutation, we create a new set of variables $n_i(t+1)$ from the current set
$n_i(t)$. Initially, all $n_i(t+1)$ are set to zero. We then repeat the
following steps $N$ times: We choose a mutation class $i$ at random for
replication, according to probabilities $p_i$ that are proportional to
$(1-s)^in_i(t)$. We then test $i$ times for backward mutations, with
probability $v$, and $L-i$ times for forward mutations, with probability $u$.
The mutation class $j$ of the new individual is then $i$ minus the successful
back mutations plus the new forward mutations. The variable $n_j(t+1)$ is
correspondingly incremented by one. After $N$ repetitions of these steps, we
have fully assembled the population of time $t+1$, and we begin again with a
new set of variables $n_i(t+2)$, and so on. The distribution at $t=0$ is
always $n_0(0)=N$, $n_i(0)=0$ for all $i>0$.

\bigskip

\noindent
\textbf{Moment expansion}

\noindent
For the case of an infinite population, we have seen that no significant
accumulation of mutations occurs if the selective strength is larger than the
per-locus deleterious mutation rate. No back mutations are necessary to
maintain this balance between selection and mutation in an infinite
population, because the wild type (with all loci carrying the '$+$' allele) is
never lost. The equilibrium point is then simply a function of the rate at
which mutants are produced from the wild type, and the rate at which the wild
type replicates in comparison to the replication rate of the mutants. The case
of a finite population is fundamentally different, because there the least
loaded class can be lost.

We have calculated the average number of mutations in a population
using a moment expansion technique previously used by
\shortciteN{WoodcockHiggs96}. In essence, we have repeated their calculations,
but have keept the forward and the backward mutation rates independent, instead
of assuming $u=v$. The details of these calculations are given in the
Appendix.

We compared our analytical result with numerical simulations. We found that
our approximations worked well for selective values $s$ of the order of $u$ or
smaller (Fig.~\ref{fig:v_khat1}), but did not give a reasonable estimate when
$s>u$. This is not too surprising, given that the calculations are done under
the assumption that $s$ is small. Unfortunately, however, the interesting
parameter region is $s>u$, for which the infinite population model allows for
a non-trivial equilibrium even in the absence of back mutations. We therefore
had to develop an alternative analytical description that would be more
suitable for $s>u$.

\bigskip

\noindent
\textbf{Effective genome size model}

\noindent
In this section, we focus on the case of no back mutations, $v=0$.
From a mathematical point of view, a finite population without back mutations
will always hit the absorbing boundary at $\langle k\rangle=L$ eventually,
i.e., it will always lose all '$+$' alleles as time
$t\rightarrow\infty$. Therefore, we know that any set of equations aimed at
finding the true equilibrium point of such a system is bound to yield
$\langle k\rangle=L$. However, this solution has no biological
significance if the waiting time to arrive at $\langle
k\rangle=L$ is extremely large, e.g., of the order of millions of generations
or larger. Consequently, we must try to find a description for the increase in
waiting times between successive clicks, and may assume that the ratchet has
stopped once the waiting times exceed a biologically relevant time scale.
Below follows a mathematical description for the waiting times that,
while being too simplistic to yield accurate quantitative results for very
small $N$, describes
well the qualitative behavior that we see in simulations, and gives valuable
insight into the nature of Muller's ratchet in finite genomes.

We assume that the number of sequences in the least loaded class $n_0$ is
given by the infinite population concentration times the population size $N$,
and that loss of the least loaded class occurs if all individuals in that
class do either fail to reproduce or give birth to mutated offspring.  The
situation of a population with a least loaded class having $k_{\min}$
mutations is equivalent to one in which the least loaded class has zero
mutations, but the genome is of length $L-k_{\min}$. In the following, we will
therefore refer to $L-k_{\min}$ as the effective genome size, and to the model
that we describe now as the ``effective genome size model''.

From \eqref{det-concs}, we find
\begin{subequations}\label{n-zero}
\begin{align}
  n_0 & = N (1-a)^{L-k_{\min}}\\\label{n-zero-exp}
  & \approx N e^{a(k_{\min}-L)}\,.
\end{align}
\end{subequations}
The probability $\xi$ that an organism in the next
generation is an offspring of one of the $n_0$ organisms in the least loaded
class is given by [with $x_i$ as defined in
\eqref{det-concs}]
\begin{equation}
  \xi = \frac{(n_0/N)(1-s)^{k_{\min}}}{\sum_{i=k_{\min}}^L x_i\, (1-s)^i}\,.
\end{equation}
The probability that there will be $j$ descendants of the least loaded class
in the next generation can be expressed in terms of $\xi$ as
\begin{equation}
  p_{\rm birth}(j) = \binom{N}{j} \xi^j(1-\xi)^{N-j}\,.
\end{equation}
The probability that all these $j$ organisms carry at least one additional
mutation is given by
\begin{equation}
  p_{\rm mut}(j) = [1-(1-u)^{L-k_{\min}}]^j\,.
\end{equation}
The total probability of a ratchet event then follows as
\begin{subequations}\label{p-ratchet}
\begin{align}
  p_{\rm ratchet}& = \sum_{j=0}^N  p_{\rm mut}(j)p_{\rm birth}(j)\notag\\
 &=  [1-\xi(1-u)^{L-k_{\min}}]^N\\
 &\approx \exp[-N\xi(1-u)^{L-k_{\min}}]\,.
\end{align}
\end{subequations}
We find that $p_{\rm ratchet}$ decays exponentially with the two
quantities $N\xi$ and $(1-u)^{L-k_{\min}}$. The latter quantity grows
exponentially with every turn of the ratchet, i.e., $(1-u)^{L-k_{\min}}$ increases exponentially
as the effective genome length $L-k_{\min}$ decreases. Moreover, $N\xi$ is
proportional to $n_0$, which again increases exponentially with decreasing
effective genome length. As a result, $p_{\rm ratchet}$ decays
superexponentially as the ratchet turns. This implies that the ratchet will
turn at a relatively high rate initially, but can suddenly reach a point where
the waiting time exceeds any biologically relevant time scale. This is
dramatically illustrated in Fig.~\ref{fig:t_kmin}. Note the logarithmic time
scale. Within the first 1000 generations, 70\% of the genes were lost, while
during the next $10^5$ generations, only an additional 10\% of the genes were
lost. Moreover, in the final 50000 generations, we observed hardly any clicks
of the ratchet at all.

Figure \ref{fig:t_kmin} gives also a comparison between simulation results and
our theoretical model. The theoretical prediction was derived from
\eqref{p-ratchet} by assuming that the average waiting time till the next
ratcheting event is given by the inverse of $p_{\rm ratchet}$, so that the
time until a certain $k_{\min}$ is reached is simply given by the sum of the
inverse ratchet probabilities from the first ratchet event until the $(k_{\rm
  min}-1)$th ratchet event. When comparing the results thus obtained to the
simulation results, we find that although the model predicts a too early
slowdown of the ratchet, the qualitative form of the predicted curve agrees
well with the one from the simulations. This demonstrates that our main
reasoning about the increase in waiting times between successive ratchet
events is correct, even though the microscopic details are somewhat incorrect.
The origin of the deviations between model and simulations is clear. At the
beginning of the simulation, we initialize all sequences to the wild type with
zero mutations, while the model always assumes a fully developed
mutation-selection equilibrium. The population in the simulation thus retains
at early times a larger fraction of individuals in the least loaded classes
than what the model assumes, and the model overestimates the transition
probabilities. For later times, the model underestimates the transition
probabilities, because it disregards cases in which the loss of the least
loaded class happens in more than one step, for example, cases in which the
number of individuals in the least loaded class first fluctuates to an
unusually low value before the currently least loaded class disappears
completely.

In Fig.~\ref{fig:u_khat}, we show another comparison between our model and the
simulation results. There, we consider the average number of mutations in the
population, $\langle k\rangle$, instead of the number of mutations in the
least loaded class, $k_{\min}$, so that we can compare our results to the
moment expansion as well. In the effective genome size model, we calculated
$\langle k\rangle$ from $k_{\min}$ with the formula $\langle k\rangle =
(L-k_{\min})a$. Both the effective genome size model and the moment expansion
agree on the mutation rate at which Muller's ratchet starts to slow down.
However, the form of the transition to a mutation free population is
significantly mis-represented by the moment expansion, while the effective
genome size model predicts the qualitative form of the transition well. In
addition, the moment expansion does not yield any insights into the temporal
change of the ratchet rate. The effective genome size model is therefore the
more useful one, despite its mathematical simplicity.

\bigskip

\noindent
\textbf{Simulation results}

\noindent
For population sizes below about one hundred, the effective
genome size model consistently predicts too low ratchet rates. This is not
surprising, as our model suffers from the same limitations that infinite
genome models do when $n_0\lesssim 1$ \shortcite{Gessler95}, namely
insufficient equilibration of the least loaded class. We therefore
have to resort to simulation results for very small population sizes.
Fig.~\ref{fig:u_khat2} shows the simulation data from Fig.~\ref{fig:u_khat}
for $N=500$, and in addition results for $N=50$ and $N=10$. All three cases
show a very similar behavior, but shifted to smaller and smaller mutation
rates. For sufficiently small mutation rates, Muller's ratchet stops after the
first few clicks. In a transition region that covers roughly one order of
magnitude in the mutation rate, Muller's ratchet stops at intermediate points,
with a good fraction of genes lost, but another large fraction of genes
retained. When the mutation rate is too high, all genes are lost within a
biologically relevant time scale.

In our analytical model presented above, we have assumed that in order to
predict the probablity of the next click of the ratchet, it is sufficient to
know the effective length $L-k_{\min}$, while the absolute values of $L$ and
$k_{\min}$ do not enter the calculation. If this assumption is correct, then
for any given mutation rate $u$, the value of $L-k_{\min}$ at which the
ratchet stops should be independent of $L$. In other words, the per-locus
mutation rate $u$ determines how many loci can at most be retained unmutated,
and the genomic mutation rate $uL$ is largely irrelevant.
Fig.~\ref{fig:max_unmutated} demonstrates that this reasoning is consistent
with our simulation results. When we plot the difference between the length
$L$ and the average number of mutations $\langle k\rangle$ as a function
of $u$, the results for widely differing lengths lie right on top of each
other in the regime in which substantial mutation accumulation occurs.

We have also performed a number of simulations with a non-zero rate of back
mutations $v$ (data not shown). The main result is similar to the case of an
infinite population, as depicted in Fig.~\ref{fig:phase_trans}. Unless the
back mutation rate is of the same order of magnitude as the forward mutation
rate, back mutations do not have a large effect on the rate at which mutations
accumulate. The ratio $u/s$ and the population size $N$ are the main
determinants of whether genomes accumulate a fair amount of mutations, or stay
largely mutation free.

\begin{center}
{\sc Discussion}
\end{center}

Two main differences between Muller's ratchet in infinite and in finite
genomes emerge from our study. First, the ratchet rate is constant for
infinite genomes, whereas the rate decays with every turn of the ratchet in a
finite genome.  The slowdown of the ratchet rate is so dramatic that the
ratchet can stop completely, even in the absence of epistasis or compensatory
mutations.  Second, in an infinite genome, the main parameter that governs the
ratchet rate, besides the population size, is the ratio $\theta$ between the
\emph{genomic} deleterious mutation rate and the selection strength
\shortcite{Haigh78}. In a finite genome, on the other hand, it is not the
genomic but the \emph{per locus} deleterious mutation rate that we have to
compare to the selection strength. As a consequence, Muller's ratchet is not
as important a limiting factor in the evolution of large genomes as was
previously thought \shortcite{MaynardSmith78}. An additional consequence is
that we can expect organisms with longer genomes and more sophisticated error
correction to be less prone to mutation accumulation than organisms with
shorter genomes, even if the mutation rates \emph{per genome} are comparable.

Backwards or compensatory mutations result in an additional slowdown of the
ratchet. However, unless they occur at a rate comparable to the forward rate,
they do not play a significant role in determining whether the ratchet stops
early, or continues until a large proportion of the genome is deteriorated. It
is therefore adequate to neglect them for order of magnitude estimates of the
ratchet rate. Nevertheless, a complete theoretical description of the ratchet
rate, including a finite genome size \emph{and} variable forward and back
mutation rates, is certainly desirable but completely lacking at this point.

The range of parameters that case Muller's ratchet to halt early is certainly
biologically plausible. For example, \shortciteN{AnderssonHughes96} estimate
the mutation rate in \emph{Salmonella typhimurium} to 0.0014--0.0072 mutations
per genome per generation, in a genome of about 200 genes~\cite{Riley93}.
\shortciteANP{AnderssonHughes96} do not estimate the selection strength $s$.
However, selection is certainly not weak, considering that out of the five
lineages in which fitness loss was observed, two experienced a doubling of
their generation time, while the other three had an increase in generation
time of about 10--15\%. Given the low mutation rate and the number of
bottlenecks of only 60, we can estimate that this loss in fitness was
probably due to only a small number of mutations, maybe between one and three.
Of course, a more accurate estimate of the selection strength in this system
would be desirable. In any case, the parameters of the simulations in
Fig.~\ref{fig:u_khat2} ($s=0.1$ and $L=100$) are probably of the right order
of magnitude, and hence we find that while bottlenecks of size
one will lead to Muller's ratchet, bottleneck sizes above 10 or 50 will not
lead to persistent genome deterioration. Therefore, if the bottlenecks
encountered during transmission from one host to another are typically between
10 to 100 individuals, bacterial
populations may not suffer significantly from mutation accumulation over
time.

In the case of RNA viruses, where mutation rates are much higher, it seems
that the risk of mutation accumulation should be higher as well.  Since
sufficiently large virus populations can exist without loss in fitness,
however, we can assume that $u/s\ll 1$, so that Muller's ratchet becomes
important only for very small populations. Most experimental work focuses on
bottleneck sizes of one
\shortcite{Chao90,Duarteetal92,Elenaetal96,Penaetal2000}, in which case loss
of fitness is readily observed after a couple of serial transfers. However,
these works do not address intermediate bottleneck sizes between 10 and 100,
or the change in fitness over time. Therefore, they do not allow a full
assessment of the importance of Muller's ratchet in virus evolution. An
important step towards more conclusive experimental results has been presented
in recent work by Chao and coworkers
\shortcite{Chaoetal97,BurchChao99,BurchChao2000}. There, a wide range of
population sizes has been investigated, and in particular in
\shortciteN{BurchChao99}, fitness measurements have been performed after every
bottleneck. Propagation of lineages of a strain with impaired fitness showed
recovery back to the original fitness level for bottleneck sizes of $N=33$ or
larger (Fig. 3 of \shortciteNP{BurchChao99}). For $N=10$, the original fitness
could not be regained within 100 generations. However,
\shortciteANP{BurchChao99} did not observe a further decline in fitness at
$N=10$ either, instead they observed a slight fitness increase.
\shortciteANP{BurchChao99}'s results for large bottlenecks demonstrate that
compensatory mutations are readily available for sufficiently large population
sizes, so that a virus population can easily recover a short sequence of
extreme bottlenecks if later it is allowed to expand again. Their results for
$N=10$ suggest two alternative interpretations. On the one hand, compensatory
and deleterious mutations may cancel each other almost exactly, so that the
net result is a small fitness increase. On the other hand, the impaired virus
strain may already have reached the point at which Muller's ratchet stops to
operate for $N=10$. In that case, the lineage is protected from further
mutation accumulation, and can safely exist until some compensatory mutations
occur eventually. We believe that the second explanation is the more accurate
one, but the current data does not allow to reject one of the two scenarios
conclusively. More data at small bottleneck sizes (between one and 10) and for
longer times should allow a more accurate assessment of these issues.

\begin{center}
{\sc Acknowledgments}
\end{center}

This work was supported by the NSF under contract No DEB-9981397. We thank
C.~Adami for carefully reading this manuscript.


\begin{center}
{\sc Appendix}
\end{center}

The following calculation is directly analogous to the one presented by
\shortciteN{WoodcockHiggs96} We refer to their work for a more complete
description of the necessary steps.

Assume that an individual $i$ carries $k_i$ deleterious mutations and creates
an offspring with $j=k_i+m_i$ deleterious mutations. The probability that an
offpring carries exactly $j$ mutations is given by the transition matrix
\begin{align}
  M_{jk} &= \sum_{i=\max(0,k-j)}^{\min(k,l-j)} \binom{k}{i}\binom{l-k}{j-k+i}
  v^iu^{j-k+i}\notag\\&\qquad\times(1-v)^{k-i}(1-u)^{l-j-i}\,.
\end{align}
The expectation values of $m_i$ and $m_i^2$ are
\begin{align}\label{A1}
  E(m_i) &= \sum_j(j-k_i)M_{jk_i} = u(l-k)-vk\,,\\
\label{A2}
E(m_i^2) &= \sum_j(j-k_i)^2 M_{jk_i} = k^2(u+v)^2\notag\\
& + k[-u^2-v^2+(v-u)+2u^2-2uL(u+v)]\notag\\
& +u(1-u)L+u^2L^2\,.
\end{align}
Both results reduce to the ones found by \shortciteANP{WoodcockHiggs96} if we
set $v=u$. Now consider two individuals $i$ and $j$ with fitnesses
$w_i=(1-s)^{k_i}$ and $w_j=(1-s)^{k_j}$ that have $n_i$ and $n_j$ offspring in
the next generation. The expected values of $n_i$, $n_i^2$, and $n_in_j$ have
been calculated by \shortciteANP{WoodcockHiggs96}, and we only state the final
result for completeness:
\begin{align}\label{A3}
  E(n_i) &\approx 1-s(k_i - \langle k \rangle)\,,\\
\label{A4}
  E(n_i^2) &\approx 2-1/N-3s(k_i-\langle k \rangle)\,,\\
\label{A5}
  E(n_in_j) &\approx 1- 1/N - s(k_i-\langle k \rangle) - s(k_j - \langle k
  \rangle)\,.
\end{align}

In what follows, we have to distinguish between averages over a population,
denoted by $\langle\dots\rangle$, and averages over an ensemble of independent
evolutionary histories, denoted by $\overline{\makebox[0cm]{\phantom{a}}\dots}\;$. The ensemble-averaged
moments of the distribution of $k_i$ are defined as
\begin{equation}
  M_n=\frac{1}{N}\overline{\sum_{i=1}^N(k_i-\langle k\rangle)^n}\,,
\end{equation}
For the second moment, we will in the following
use the symbol $V$ instead of $M_2$.

Now, write
\begin{equation}
  \overline{\langle k\rangle}_{t+1}=\frac{1}{N}\overline{\sum_i
  n_i(k_i+m_i)}\,.
\end{equation}
With \eqref{A1} and \eqref{A3}, and the equilibrium condition
$\overline{\langle k\rangle}_{t+1} = \overline{\langle k\rangle}_{t+1} =
\overline{\langle k\rangle}$, we find
\begin{equation}\label{A6}
(u+v)\overline{\langle k\rangle}+[1-(u+v)]sV = uL\,.
\end{equation}
Similarly, from
\begin{equation}
  \overline{\langle k^2\rangle}_{t+1}=\frac{1}{N}\overline{\sum_i
  n_i(k_i+m_i)^2}
\end{equation}
and
\begin{equation}
  \overline{\langle k\rangle^2}_{t+1}=\frac{1}{N^2}\overline{\sum_i\sum_j
  n_in_j(k_i+m_i)(k_j+m_j)}
\end{equation}
we find in equilibrium, using \eqref{A1}--\eqref{A5},
\begin{align}\label{A7}
  V &= [1-(u+v)]^2[(1-1/N)V-sM_3]\notag\\&+[u(u-1)-v(v-1)]
\times[(1-1/N)
  \overline{\langle k\rangle} -sV]\notag\\&+u(1-u)L(1-1/N)\,.
\end{align}
These two equations contain three variables, and in general every equation of
higher order will contain correspondingly higher moments, so that the set of
equations can never be solved. This problem can be avoided with a
suitable closure approximation. Following \shortciteANP{WoodcockHiggs96}, we
use
\begin{equation}\label{A8}
  M_3 = V(1-2\overline{\langle k\rangle}/L),
\end{equation}
which is the expression for the third moment in an infinite population.
With \eqref{A6}, \eqref{A7}, and \eqref{A8}, $\overline{\langle k\rangle}$ is
uniquely determined, and we find
\begin{equation}\label{A9}
 \frac{\overline{\langle k\rangle}}{L} = \frac{\alpha}{4}-\frac{1}{2}\sqrt{\Big(\frac{\alpha}{2}\Big)^2-\beta}\,,
\end{equation}
with
\begin{align}
  \alpha &= \frac{3u+v-2}{u+v-1} + \frac{1}{sN} +
  \frac{2(u+v)-(u+v)^2}{s(u+v-1)^2} + \frac{u-v}{N(u+v)}\,,\\
  \beta &= \frac{2u}{u+v}\Big[\frac{2(u+v)-(u+v)^2}{s(u+v-1)^2} + \frac{1}{sN}
  + \frac{u}{u+v-1}\notag\\&\qquad + \frac{1}{N} \frac{u-1}{u+v-1}\Big]
\end{align}

\newpage

\begin{figure}
\caption{\label{fig:phase_trans}Average fraction of mutations in an infinite
  population as a function of $u/s$ for various back mutation rates $v$.
}
\end{figure}

\begin{figure}[p]
\caption{\label{fig:v_khat1}Average number of mutations $\langle k\rangle$
  vs.\ the back mutation rate $v$, for $N=100$, and $s=u=0.01$.
  Solid lines represent the analytic
  prediction based on \eqref{p-ratchet}, points represent simulation results,
  averaged over 5
  replicates. The errors in the simulation  are of the order of
  the symbol size.
}
\end{figure}

\begin{figure}[p]
\caption{\label{fig:t_kmin}Number of mutations in the least loaded class as a
  function of time, for $N=1000$, $L=100$, $s=0.1$, $u=0.01$, and $v=0$. The
  thin solid lines represent simulation results, the thick dashed line
  represents the theoretical prediction obtained from \eqref{p-ratchet}.}
\end{figure}

\begin{figure}[p]
\caption{\label{fig:u_khat}Average number of mutations in the population
  vs.\ genomic mutation rate, for $N=500$, $L=100$, $s=0.1$, and $v=0$. The
  individual points represent results from simulation results, averaged over 5
  replicates. The errors are of the order of the symbol size. The solid line
  represents the theoretical prediction obtained from \eqref{p-ratchet}, and
  the dashed line represents the moment expansion result \eqref{A9}.  The
  simulation results and equation \eqref{p-ratchet} were evaluated after
  $t=10^5$ generations. The moment expansion predicts an equilibrium point, so
  time does not enter this equation.}
\end{figure}

\begin{figure}[p]
\caption{\label{fig:u_khat2}Average number of mutations in the population
  vs.\ genomic mutation rate, for various population sizes $N$, and $L=100$,
  $s=0.1$, $v=0$. The individual points represent results from simulation
  results, averaged over 5 replicates. The errors are of the order of the
  symbol size.  }
\end{figure}

\begin{figure}[p]
\caption{\label{fig:max_unmutated}Average number of mutation-free loci
  $L-\langle k \rangle$ vs.\ mutation rate $u$, for various length $L$
  and $N=50$, $s=0.1$, $v=0$.  The individual points represent results from
  simulation results, averaged over 5 replicates. The errors are of the order
  of the symbol size.  }
\end{figure}

\clearpage
\onecolumn

\noindent Figure 1:

\centerline{
\includegraphics[width=18cm,angle=90]{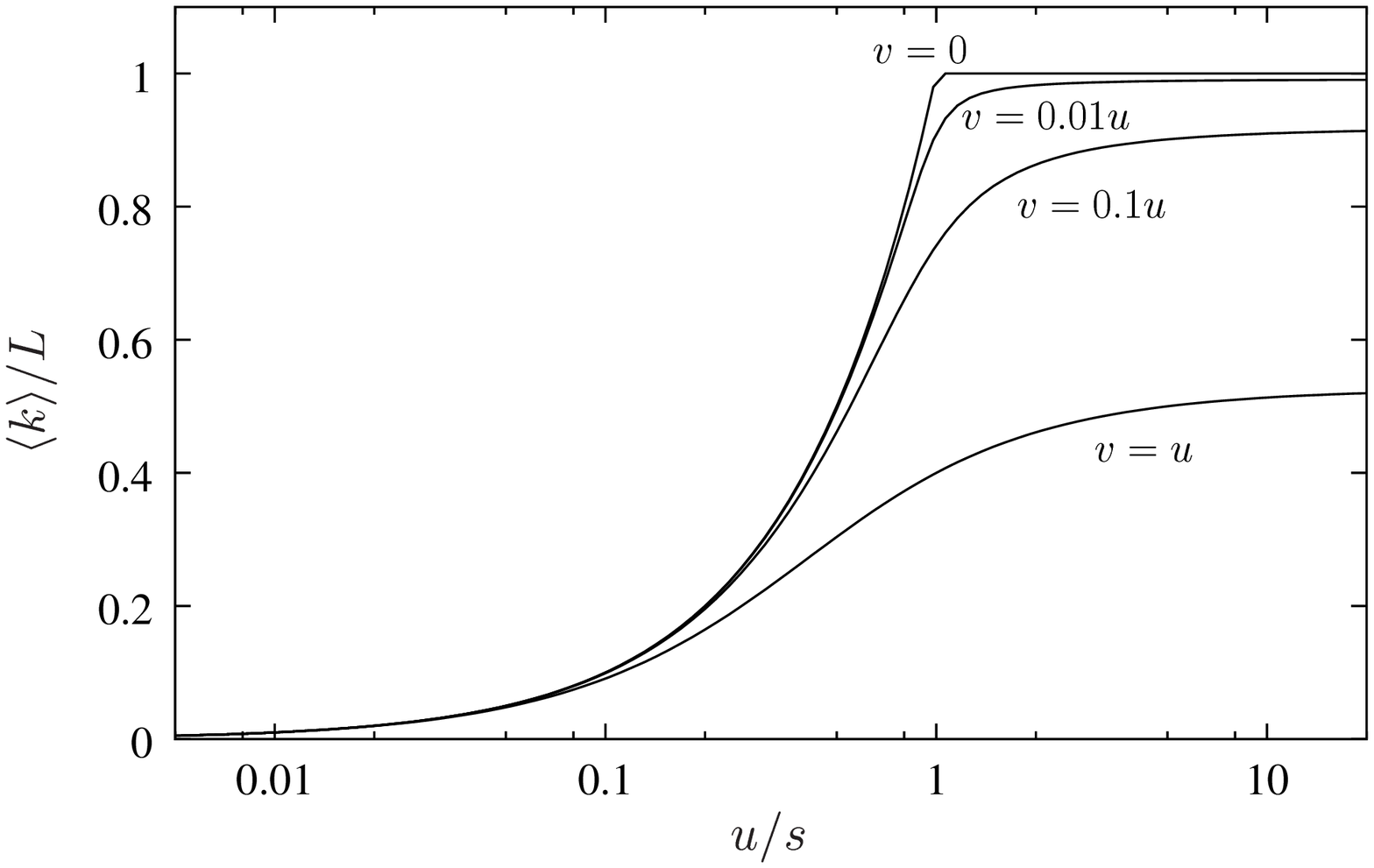}
}

\newpage
\noindent Figure 2:

\centerline{
\includegraphics[width=18cm,angle=90]{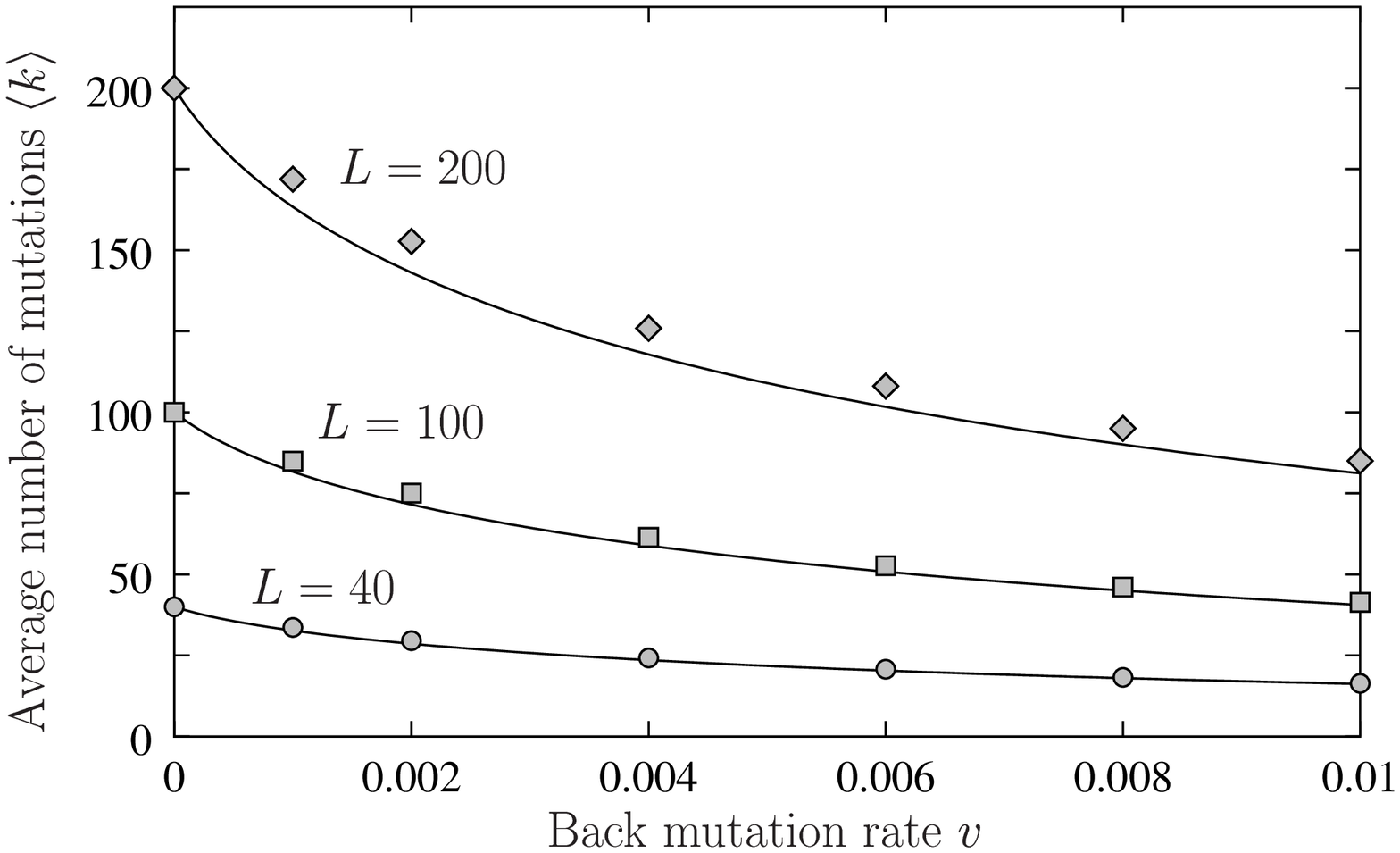}
}

\newpage
\noindent Figure 3:

\centerline{
\includegraphics[width=18cm,angle=90]{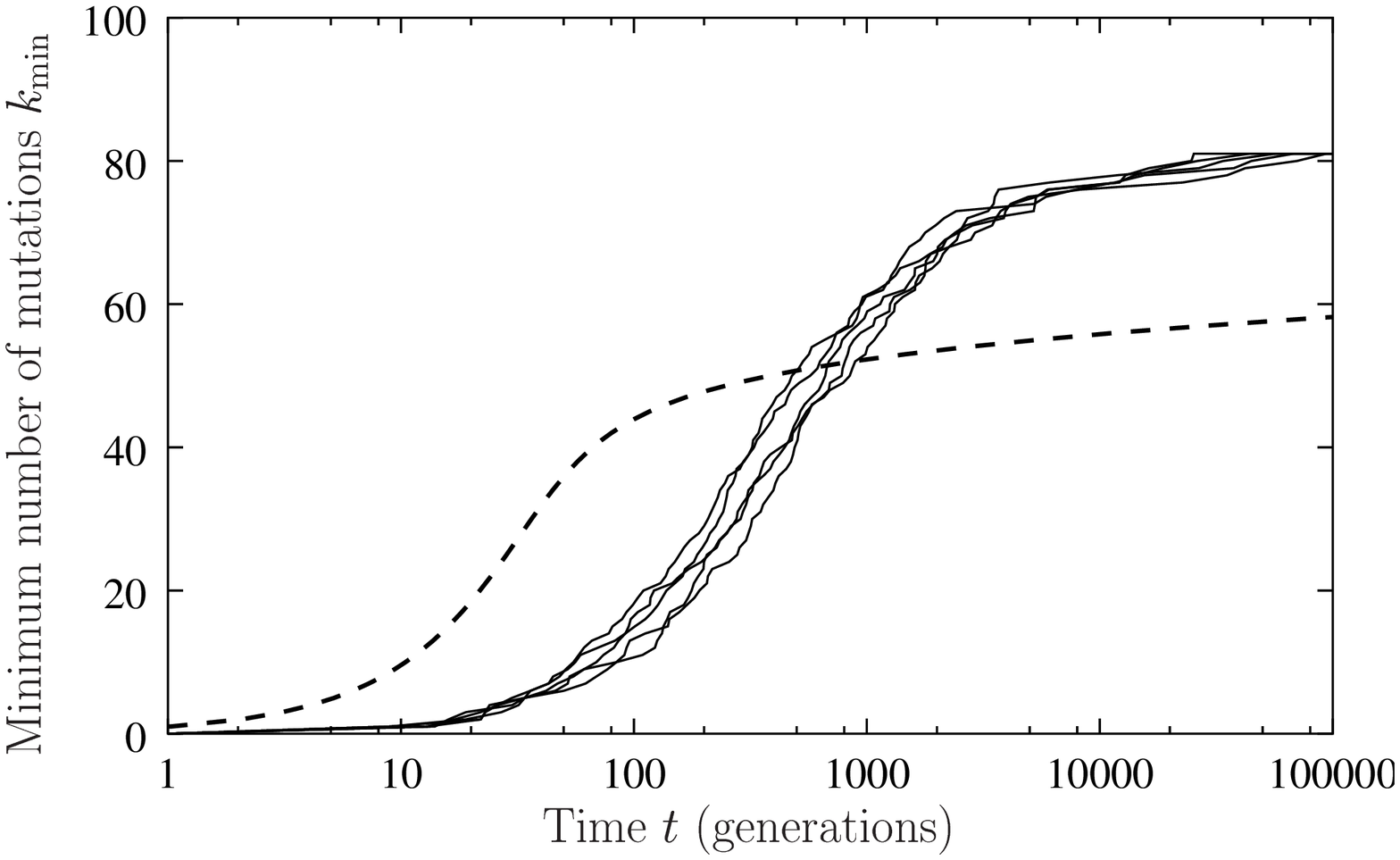}
}

\newpage
\noindent Figure 4:

\centerline{
\includegraphics[width=18cm,angle=90]{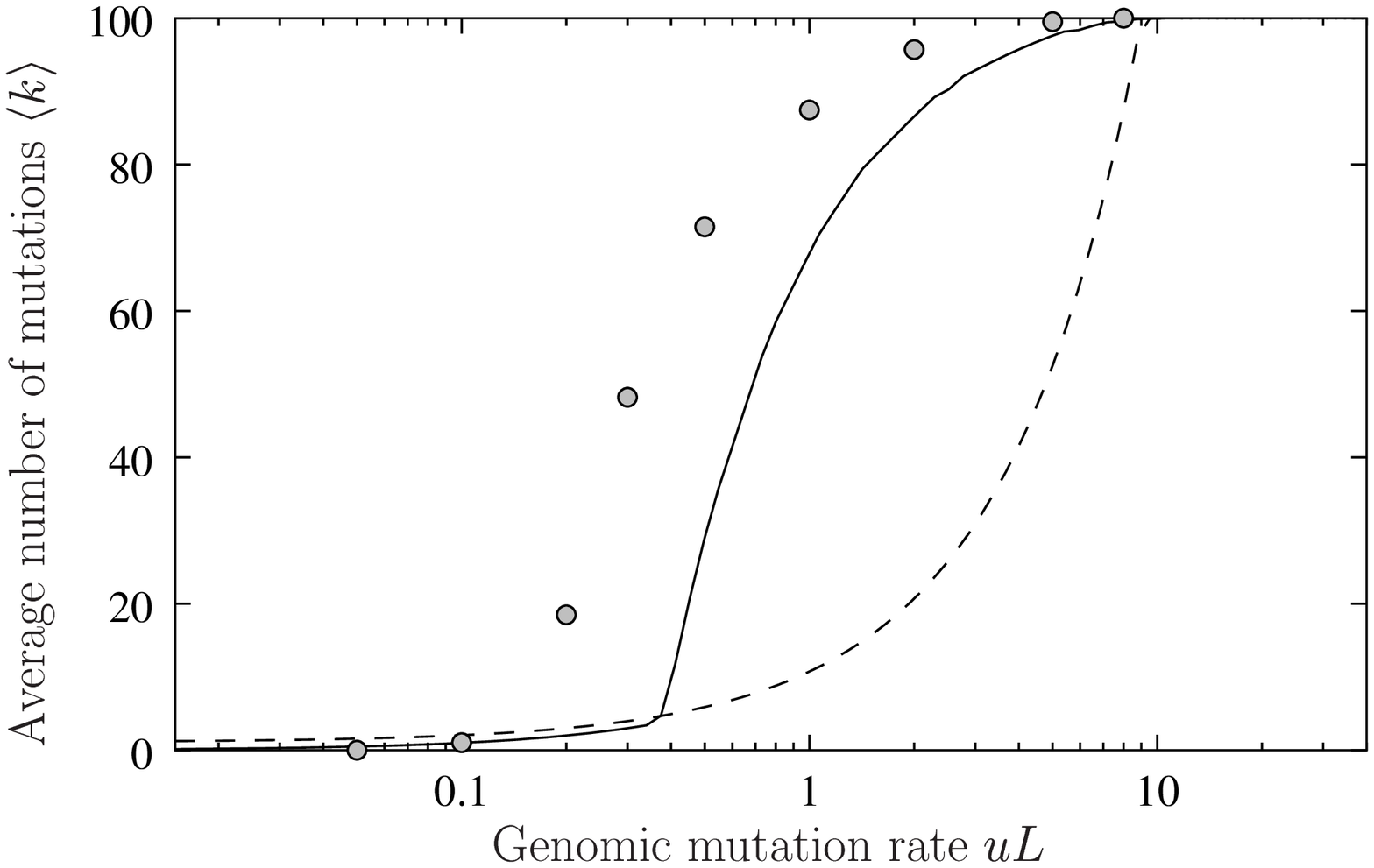}
}

\newpage
\noindent Figure 5:

\centerline{
\includegraphics[width=18cm,angle=90]{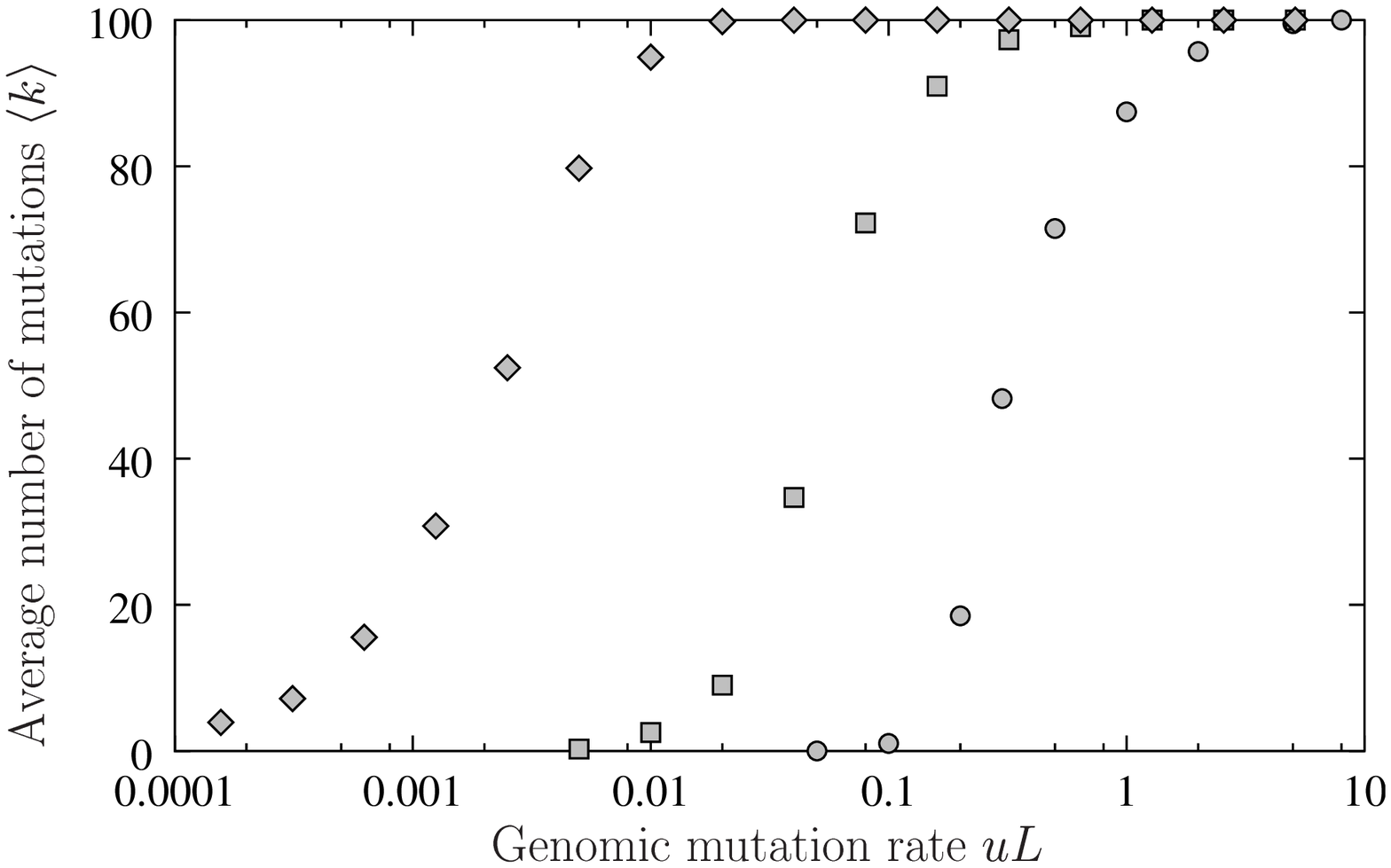}
}

\newpage
\noindent Figure 6:

\centerline{
\includegraphics[width=18cm,angle=90]{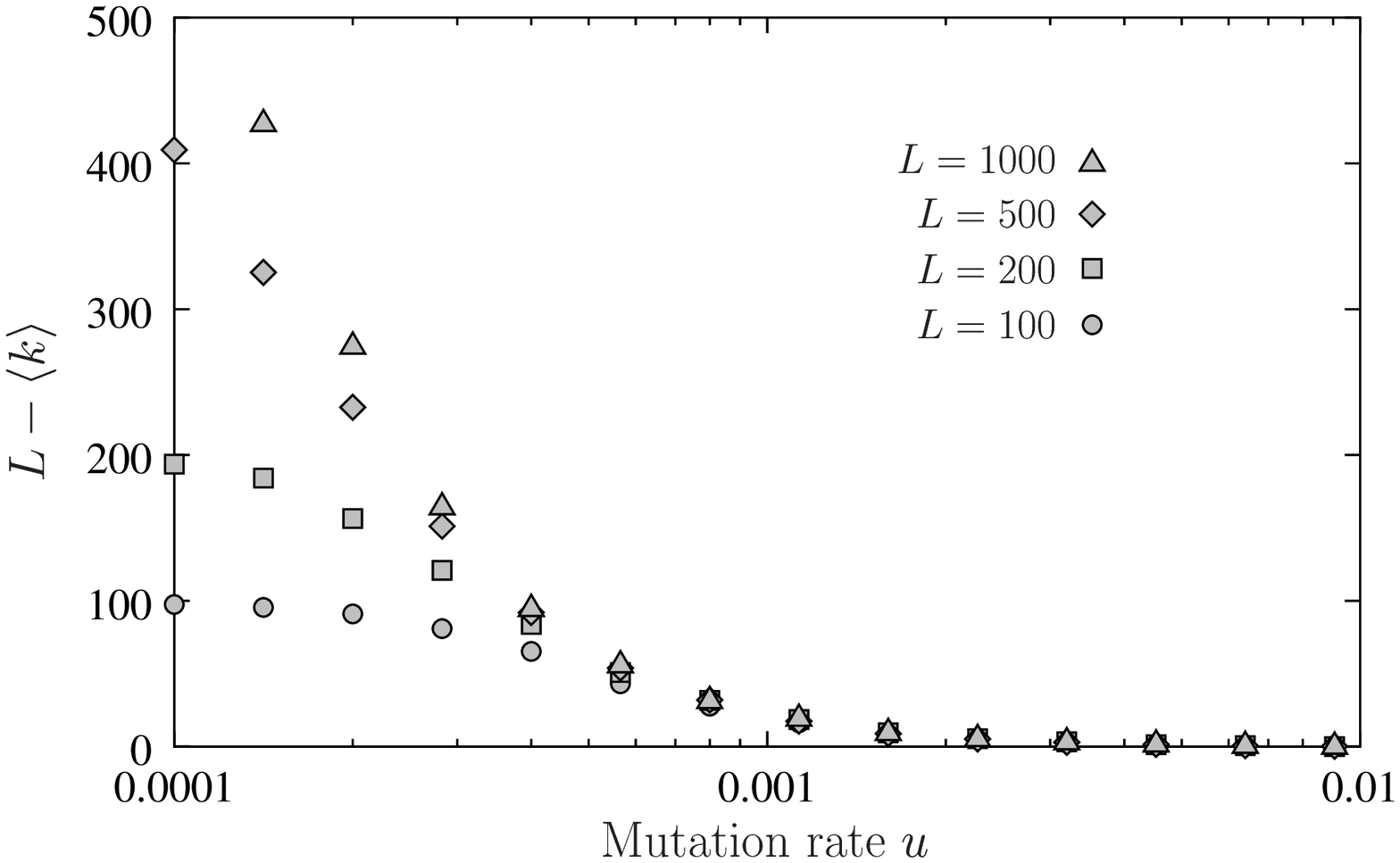}
}

\end{document}